\documentclass[3p]{elsarticle}
\usepackage{lineno,hyperref}
\modulolinenumbers[5]
\journal{Simpósio Brasileiro de Automação Inteligente 2023}

\usepackage[english,brazil]{babel}
\usepackage[T1]{fontenc}
\usepackage[utf8]{inputenc}
\usepackage{ae}
\usepackage{graphicx}
\usepackage{amssymb,amsmath}
\usepackage{enumitem}
\usepackage{mathtools}
\usepackage{placeins}
\usepackage{algorithm,algorithmic}
\usepackage{color}
\usepackage[dvipsnames]{xcolor}
\usepackage{array}

\newtheorem{theorem}{Teorema}

\newtheorem{assumption}{Hipótese}

\newtheorem{proposition}{Proposição}
\newtheorem{remark}{Nota}
\newtheorem{example}{Exemplo}
\newtheorem{tutorialstep}{Passo do Tutorial}

\newtheorem{proofbreve}{Esboço de prova}

\begin{document}
\selectlanguage{english}	
	
\begin{frontmatter}
                       
\title{Tutorial: Implementando Controladores Preditivos Não Lineares através do Ferramental LPV}


\author[adufsc1,adgipsa]{Marcelo M. Morato}
\author[adufsc1]{Amir Naspolini}
\author[adufsc1]{Julio E. Normey-Rico}

\address[adufsc1]{Dept. de Automa\c{c}\~ao e Sistemas
  (\emph{DAS}), Univ. Fed. de Santa Catarina,
  Florian\'opolis-SC, Brazil.}

\address[adgipsa]{Univ. Grenoble Alpes, CNRS, Grenoble INP (Institute of Engineering Univ. Grenoble Alpes), GIPSA-lab, 38000 Grenoble, France. 
  (marcelomnzm@gmail.com)}

\renewcommand{\abstractname}{{\bf Abstract:~}}

\begin{abstract} 
Recent works have demonstrated how Linear Parameter Varying Model Predictive Control (LPV MPC) algorithms are able to control nonlinear systems with precision and reduced computational load. Specifically, these schemes achieve comparable performances to state-of-the-art nonlinear MPCs, while requiring the solution of only one quadratic programming problem (thus being real-time capable). In this tutorial paper, we provide a step-by-step overview of how to implement such LPV MPC algorithms, covering from modelling to stability aspects. For illustration purposes, we consider a realistic implementation for a  gas-lift petroleoum extraction process, comparing the LPV approach with the becnhmark nonlinear MPC software CasADi.
 
\vskip 1mm
\selectlanguage{brazil}
{\noindent \bf Resumo}\\ 
Diversos estudos recentes demonstram como Controladores Preditivos baseados em Modelo Lineares a Parâmetros Variantes (MPC LPV) têm a capacidade de regular sistemas não lineares com precisão e esforço numérico reduzido. Tais controladores podem alcançar desempenhos comparáveis com abordagens do tipo MPC não linear, ao passo que requerem apenas a solução de um problema de programação quadrática por amostra (possibilitando aplicações embarcadas, em tempo-real). Neste artigo-tutorial, detalhamos todas as etapas necessárias para o projeto de algoritmos MPC LPV; abordamos temas referentes à modelagem (como obter reapresentações LPV), às hipóteses necessárias, e às garantias de estabilidade, além de debatermos noções sobre rastreamento de referências e predição dos sinais de agendamento LPV. Para fins de ilustração, apresentamos uma implementação numérica para um sistema de extração de petróleo por elevação artificial com injeção de gás (\textit{gas-lift}),
comparando a abordagem LPV com o software MPC não linear de ponta CasADi.
\end{abstract}

\selectlanguage{english}

\end{frontmatter}

\selectlanguage{brazil}	
\section{Introdução}
\label{intro}

O controle preditivo baseado em modelo (MPC) é amplamente difundido para a regulação ótima de sistemas sujeitos a restrições \citep{camacho2013model}. A ideia central concentra-se na obtenção da lei de controle através da solução, a cada amostra, de um problema de otimização levando em conta as restrições do processo e \emph{predições} do comportamento futuro das variáveis de interesse ao longo de um \emph{horizonte deslizante de predição}. Ao longo das últimas duas décadas, desenvolveu-se um arcabouço teórico extenso para esta metodologia de controle, incluindo \emph{certificados} de performance e estabilidade (em malha fechada), tanto para processos lineares quanto não lineares, c.f.  \citep{limon2018nonlinear,santos2019constraint, morato2023IQCstabilizing}. Entretanto, o grande desafio teórico enfrentado pela comunidade científica é: como possibilitar a aplicação de algoritmos MPC \emph{não lineares} (NMPC) em \emph{tempo-real}, na escala dos microssegundos \citep{quirynen2015autogenerating}. O impedimento para tal se dá pelo fato de que, quando usamos modelo não linear, o programa de otimização do MPC resultante torna-se, também, não linear, requerendo um esforço numérico elevado para resolução. Enfatizamos, tal como \citet{koehler2019nonlinear}, que algoritmos NMPC apresentam um crescimento exponencial do tempo médio de cômputo face à dimensão do sistema e ao tamanho do horizonte predito.

A solução mais difundida para este problema é a aproximação do programa de otimização não linear por uma sequência de programas de programação quadráticos (QPs), assim reduzindo a carga numérica\footnote{Ressaltamos que QPs podem ser resolvidos online com a vasta maioria de \emph{solvers} disponíveis na atualidade, permitindo, assim, com que a lei de controle preditiva seja gerada em tempo-real (embarcada). }. Este tópico é proficuamente discutido por \citet{gros2020linear}, que detalha como o estado-da-arte sobre NMPCs \emph{rápidos} consiste, em aproximações por iterações em tempo-real e gradientes.

Apesar da profusão recente de ferramentas como ACADO e CasADi \cite{andersson2019casadi}, debatemos, aqui, um caminho alternativo: o projeto através de modelos Lineares a Parâmetros Variantes (LPV), cujos principais trabalhos são listados na revisão sistemática \citep{morato2020model}. A grande vantagem do uso de modelos LPV\footnote{Em suma, modelos LPV são capazes de representar trajetórias não lineares, variantes no tempo, através do uso de parâmetros de agendamento (variantes no tempo) limitados e mensuráveis online.} consiste no fato da otimização resultante ser do tipo QP. Desta forma, mantém-se uma carga numérica reduzida, e uma predição (quase) \textbf{exata} das dinâmicas futuras do sistema controlado, uma vez que nenhuma etapa de aproximação é necessária. Recentemente, diversos algoritmos MPC LPV foram  desenvolvidos, considerando, por exemplo, rastreamento de referências \citep{cisneros2017fast}, estabilização robusta \citep{morato2021robustzono}, e validações experimentais \citep{calderon2019qlpv,alcala2020autonomous}. Em comparação com os algoritmos NMPC rápidos, diferentes resultados apontam para \textbf{competitividade numérica} e \textbf{desempenho similare}, e.g. \citep{cisneros2019wide,morato2022nmpccassestudy}.

Com base neste contexto, a contribuição deste artigo é de servir como um \textbf{tutorial} detalhado, passo-a-passo, para o projeto de algoritmos MPC rápidos para sistemas não lineares, com base em modelos LPV. Na sequência, abordamos os seguintes tópicos: obtenção de modelos LPV para sistemas não lineares (Seção \ref{sec2}); predição dos parâmetros de agendamento, com base em expansões de Taylor (Seção \ref{sec3}); implementação (Seção \ref{sec4}); garantias de estabilidade através de ingredientes terminais (Seção \ref{sec5}); rastreamento de referências (Seção \ref{sec6}); ensaios em simulação (Seção \ref{sec7}); conclusões (Seção \ref{secconc}).

\noindent \textbf{Notação.} O conjunto de índices $\mathbb{N}_{[a , b]}$
é dado por $\{i \, \in \, \mathbb{N} \, | \, a \, \leq \, i \, \leq \,
b\}$, com $0 \, \leq \, a \, \leq  \, b$. A matriz identidade de tamanho $j$ é representada por $I_{j}$, ao passo que $I_{\{i\},j}$ representa a $i$-ésima linha da respectiva matriz de identidade. col$(v)$ representa a vetorização em coluna das entradas $v$; $u(k+i|k)$ representa a predição da variável $u$ para o instante de tempo $k+i$, realizada no instante $k$. Funções classe $\mathcal{K}$ são escalares, positivas e estritamente crescentes.

\subsection*{Estudo de caso: Extração de Petróleo por \text{gas-lift}}
\label{motivacaocasodeestudo}

\noindent Para ilustrarmos e debatermos as técnicas descritas ao longo deste tutorial, consideramos um estudo de caso concreto:  o processo de extração de petróleo do fundo de um poço por elevação artificial, através da injeção de gás, método chamado de \textit{gas-lift}. Este sistema é descrito a seguir.

\begin{figure}
\centering
\includegraphics[width=1\linewidth]{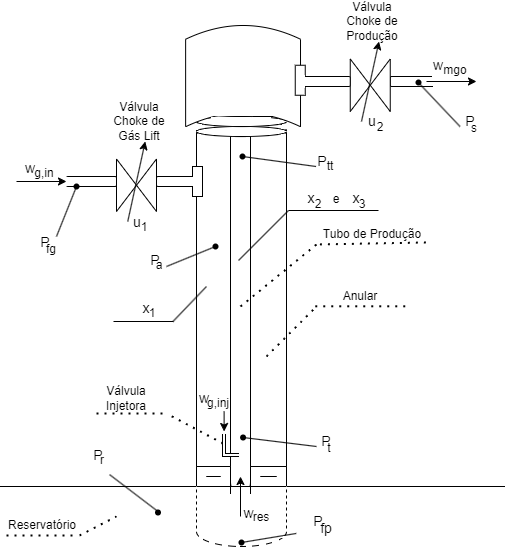}
\caption{\textit{Gas-lift} \citep{plucenio2009gas}.}
\label{GLimage}
\end{figure}

Em geral, para que se possa extrair petróleo de um reservatório e conduzi-lo até a superfície, é necessário usar alguma estratégia de elevação artificial no poço produtor. A técnica do \textit{gas-lift}, tal como ilustrado na Figura \ref{GLimage}, consiste na injeção de um gás pressurizado oriundo da plataforma até o fundo do mar através de uma tubulação exterior ao poço (tubo anular). Esta injeção é feita através de válvulas ao longo do tubo de produção, de forma a diminuir a  
densidade do fluido que está sendo extraído e permitir que a pressão do reservatório seja suficiente para elevá-lo até a superfície. O gás usado para este processo é, em geral, extraído do próprio reservatório e seu processamento se dá em unidades instaladas na plataforma.

O processo de \textit{gas-lift} é descrito matematicamente pelo modelo apresentado a seguir, considerando pressão uniforme ao longo do anular e que a mistura gás-líquido é descrita unicamente em função do óleo e do gás \citep{plucenio2009gas}:
\begin{eqnarray}   
\label{EDOsMassa}
\left\{\begin{array}{rcl}
\dot{x}_1(t)&=&w_{g,in}(t)-w_{g,inj}(t) \, \text{,}\\ 
\dot{x}_2(t)&=&w_{g,inj}(t)+w_{g,res}(t)-w_{g,out}(t)  \, \text{,} \\ 
\dot{x}_3(t)&=&w_{o,res}(t)-w_{o,out}(t)  \, \text{,}\end{array}\right.
\end{eqnarray}
\noindent sendo $x_1$ é a quantidade de gás no anular, $x_2$  e $x_3$ a quantidade de gás e óleo no tubo, respectivamente. As vazões mássicas referentes ao gás $w_{g,in}$, $w_{g,inj}$, $w_{g,res}$ e $w_{g,out}$ são, respectivamente, a vazão de entrada no anular, a vazão entre o anular e o tubo, a vazão entre o reservatório e o tubo e a vazão de saída do tubo. As vazões mássicas referentes ao óleo $w_{o,res}$ e $w_{o,out}$ são, respectivamente, as vazões de saída do reservatório e do tubo. A vazão mássica total proveniente do reservatório, $w_{res}$, é composta pela vazão de óleo $w_{o,res}$ e de gás $w_{g,res}$. Especificamente, temos:
\begin{eqnarray}   
\nonumber
w_{g,in}(t)&=&K_{cgl}\sqrt{\max(0,\mu_a(t)(P_{fg}-P_a(t)))}u_1(t) \, \text{,}\\ \nonumber
w_{g,inj}(t)&=&K_{inj}\sqrt{\max(0,\mu_a(t)(P_a(t)-P_t(t)))} \, \text{,}\\ \nonumber
w_{o,res}(t)&=&i_P\max(0,(P_r-P_{fp}(t))) \, \text{,}\\ \nonumber
w_{g,res}(t)&=&g_{go}w_{o,res}(t) \, \text{,}\\ \nonumber
w_{mgo}(t)&=&K_{cp}\sqrt{\max(0,\mu_t(t)(P_{tt}(t)-P_s))}u_2(t) \, \text{,}\\ \nonumber
w_{g,out}(t)&=&x_2(t)(x_2(t)+x_3(t))^{-1}w_{mgo}(t) \, \text{,}\\ \nonumber
w_{o,out}(t)&=&x_3(t)(x_2(t)+x_3(t))^{-1}w_{mgo}(t) \, \text{,}
\end{eqnarray}
\noindent sendo $w_{mgo}$ a vazão da mistura gás-líquido na saída do tubo, $u_1$ e $u_2$ os sinais de controles correspondentes às aberturas das válvulas (\textit{choke}) de entrada de gás e de produção de óleo. $\mu_a$ e $\mu_t$ representam as densidades de gás no anular e da mistura gás-líquido, respectivamente;  $P_a$, $P_t$, $P_{fp}$, $P_{tt}$ representam, respectivamente, as pressões no anular, no tubo, no fundo do poço e no topo do tubo. Essas variáveis são descritas através das seguintes relações:
\begin{eqnarray}   
\nonumber
P_a(t)&=&\left(\frac{T_aR}{M_gV_a}+\frac{gL_a}{H_a}\right)x_1(t)\,\text{,}\\ \nonumber
P_{tt}(t)&=&T_tR\mu_o x_2(t)M_g^{-1} \left(V_t\mu_o-x_3(t)\right)^{-1}\,\text{,}\\ \nonumber
P_t(t)&=&P_{tt}(t)+(x_2(t)+x_3(t))gA_t^{-1}\,\text{,}\\ \nonumber
P_{fp}(t)&=&P_t(t)+\mu_ogH_{fp}\,\text{,}\\ \nonumber
\mu_a(t)&=&M_g(T_aR)^{-1}P_a(t)\,\text{,}\\ \nonumber
\mu_t(t)&=&(x_2(t)+x_3(t))(V_t)^{-1} \,\text{.}
\end{eqnarray}

\begin{remark}
Todos os parâmetros são descritos na Tabela \ref{parametrização}.
\end{remark}
\begin{table}[h!]
    \centering
      \caption{Parâmetros do sistema de \textit{gas-lift}.}
    \label{parametrização}
    \small
    \begin{tabular}{c >{\centering}b{3cm} c c}
        \hline
         Parâmetro & Descrição &  Unidade & Valor \\\hline 
         $K_{cgl} $ & \parbox{3cm}{Constante da válvula choke de Gás Lift} & $\sqrt{kgm^3Pa^{-1}}s^{-1}$  &2e-3\\ \hline
         $K_{inj} $ &  \parbox{3cm}{Constante da válvula Injetora}  & $\sqrt{kgm^3Pa^{-1}}s^{-1}$  &1e-4\\\hline
         $i_P $ & \parbox{3.3cm}{Índice de Produção}  & $kgPa^{-1}s^{-1}$  &2.6e-6\\\hline
         $K_{cp} $ & \parbox{3cm}{Constante da válvula choke de produção} & $\sqrt{kgm^3Pa^{-1}}s^{-1}$ &2e-3\\\hline
         $P_{fg} $ & \parbox{3cm}{Pressão da fonte de gás}  & bar &191\\\hline
         $P_{r} $ & \parbox{3cm}{Pressão no reservatório}  & bar   &150\\\hline
         $P_{s} $ & \parbox{3cm}{Pressão no separador}  & bar   &20\\\hline
         $g_{go}$ & \parbox{3cm}{Proporção Gás-Oléo}  & -   &0.01\\\hline
         $T_{a} $ & \parbox{3cm}{Temperatura no anular}  & K   &301\\\hline
         $T_{t} $ & \parbox{3cm}{Temperatura no tubo}  & K   &305\\\hline
         $R$ & \parbox{3cm}{Constante universal dos gases}  & Jmol$^{-1}$K$^{-1}$   &8.31\\\hline
         $M_{g} $ &\parbox{3cm}{Massa molar do gás}  & kgmol$^{-1}$ &0.028\\\hline
         $V_{a} $ & \parbox{3cm}{Volume do anular}  & m$^{3}$   &24.8343\\\hline
         $H_{a} $ & \parbox{3cm}{Altura do anular}  & m   & 1500\\\hline
         $V_{t} $ &\parbox{3cm}{Volume do tubo}  &  m$^{3}$   & 17.2485\\\hline
         $H_{t} $ & \parbox{3cm}{Altura do tubo } & m   & 1500\\\hline
         $A_{t} $ & \parbox{3cm}{Área do tubo} & m$^2$   & 0.0115\\\hline
         $H_{fp} $ & \parbox{3cm}{Altura do fundo do poço } & m   & 500\\\hline
         $\mu_{o} $&\parbox{3cm}{Densidade do óleo} & kgm$^{-3}$   & 900\\
         \hline 
    \end{tabular}
\end{table}

\section{Representações LPV}
\label{sec2}

\noindent A primeira etapa para a síntese de controladores preditivos não lineares tipo LPV consiste na obtenção de uma representação LPV exata do processo não linear controlado, descrito pela seguinte equação à diferenças:
\begin{eqnarray}
\label{sisnonlin}
x(k+1) &=& f\left(x(k),u(k)\right) \, \text{,} \\ \nonumber
y(k) &=& g\left(x(k),u(k)\right) \, \text{.}
\end{eqnarray}

Na Eq. \eqref{sisnonlin}, consideramos que as funções (não lineares) $f(\cdot)$ e $g(\cdot)$ são analíticas, contínuas, diferenciáveis e possuem todas as derivadas contínuas e limitadas. No estudo de caso do processo de \textit{gas-lift}, a função $f(\cdot)$ é obtida diretamente pela discretização Euler da Eq. \eqref{EDOsMassa}. 

Consideramos que os estados $x \, \in \, \mathbb{R}^{n_x}$ são mensuráveis para toda amostra discreta $k\,\geq\,0$, tal como as saídas $y \, \in \, \mathbb{R}^{n_y}$. No contexto de MPC, $u \, \in \, \mathbb{R}^{n_u}$ define um sinal de controle preditivo que deve garantir um desempenho ótimo (discussões na Seção \ref{sec4}), ao passo que deve satisfazer as \emph{condições de admissibilidade} do sistema, ou seja, um regulação condizente com as restrições físicas e naturais do processo, representadas matematicamente por:
\begin{eqnarray}\label{XUYcalconstraints}
   \left\{\begin{array}{ccc} x(k) &\in& \mathcal{X}, \forall k \, \geq \, 0\, \text{,} \\ 
    y(k) &\in& \mathcal{Y}, \forall k \, \geq \, 0\,\text{,} \\ 
    u(k) &\in& \mathcal{U}, \forall k \, \geq \, 0 \, \text{,} \end{array}\right.
\end{eqnarray}
\noindent sendo $\mathcal{X}$, $\mathcal{Y}$, $\mathcal{U}$ conjuntos fechados, convexos e conhecidos. Em termos práticos, estes conjuntos definem condições derivadas da prática; por exemplo, uma válvula de injeção de gás, cuja abertura é limitada entre $0$ a $100\,\%$, define um conjunto de controle admissível $\mathcal{U}$ do tipo $[0 \, , 1]$. Para fins de praticidade, consideramos doravante os seguintes conjuntos tipo caixa, considerando $\overline{v}_i$ como o limite de magnitude para a $i$-ésima componente da variável $v$:
\begin{eqnarray}\label{XUYcalconstraints2}
\left\{\begin{array}{ccc}
   \mathcal{X} &:=& \left\{ x \in \mathbb{R}^{n_x}\, | \, |x_j|\,\leq\,\overline{x}_j \, \forall j \in \mathbb{N}_{[1,n_x]}\right\} \, \text{,} \\
   \mathcal{Y} &:=& \left\{ y \in \mathbb{R}^{n_y}\, | \, |y_j|\,\leq\,\overline{y}_j \, \forall j \in \mathbb{N}_{[1,n_y]}\right\} \, \text{,} \\
   \mathcal{U} &:=& \left\{ u \in \mathbb{R}^{n_u}\, | \, |u_j|\,\leq\,\overline{u}_j \, \forall j \in \mathbb{N}_{[1,n_u]}\right\} \, \text{.} \end{array}\right.
\end{eqnarray}

Uma vez conhecida a representação discreta não linear do processo controlado, Eq. \eqref{sisnonlin}, tal como as restrições de admissibilidade, Eq. \eqref{XUYcalconstraints}, o \textbf{primeiro passo} necessário é a transcrição das trajetórias do sistema, dentro dos domínios admissíveis, através de modelo LPV. Para tal, uma série de abordagens pode ser considerada, tal como identificação LPV global, e.g. \citep{piga2015lpv}, e inclusões diferenciais lineares, convexas e côncavo-convexas, e.g. \citep{shamma2012overview}. 
Devido a sua simplicidade e notório sucesso, c.f. \citep{hoffmann2014survey, morato2020model,robles2019performance}, consideramos o uso das chamadas inclusões quasi-LPV exatas (\emph{qLPV embeddings}). Para tal, pressupomos que, no contexto do sistema controlado, há como encapsular as não linearidades em variáveis de agendamento mensuráveis. Analiticamente, tal hipótese impõe a necessidade de existência de mapas analíticos, limitados e diferenciáveis $h_1(x,u) \, \in \, \mathbb{R}^{(n_x+n_y)\times n_x}$ e $h_2(x,u) \, \in \, \mathbb{R}^{(n_x+n_y)\times n_u}$ de tal forma que, para todo $x \, \in \, \mathcal{X}$ e todo $u \, \in \, \mathcal{U}$, verifique-se:
\begin{eqnarray} \label{embeddingNC}
    \left[\begin{array}{c}f(x,u) \\ g(x,u) \end{array}\right] &=& h_1(x,u)x + h_2(x,u)u \, \text{.} 
\end{eqnarray}

Assim, com base na condição de necessariedade imposta pela Eq. \eqref{embeddingNC}, o modelo original da Eq. \eqref{sisnonlin} pode ser rescrito na seguinte forma LPV:
\begin{eqnarray} \label{modeloLPV}
    x(k+1)&=& A(\rho(k))x(k) + B(\rho(k))u(k) \, \text{,} \\ \nonumber
    y(k) &=& C(\rho(k))x(k) + D(\rho(k))u(k) \, \text{.}
\end{eqnarray}

No modelo LPV da Eq. \eqref{modeloLPV}, as variáveis de agendamento são, por definição, limitadas, ou seja, $\rho(k) \, \in \, \mathcal{P}, \, \forall \, k \geq 0$ e, ademais, é imposta uma relação de interdependência com os estados e os sinais de controle, ou seja:
\begin{eqnarray} \label{modeloLPV2}
    \rho(k) &=& f_\rho\left(x(k),u(k)\right) \, \text{.}
\end{eqnarray}

A função $f_\rho(x,u)$ é chamada de \emph{proxy LPV}, ao passo que o conjunto $\mathcal{P}$ é chamado de conjunto de agendamento. Em termos práticos, ressaltamos que, uma vez que os sinais $x$ e $u$ são conhecidos e mensuráveis, o parâmetro $\rho$ pode ser diretamente inferido. A seguir, apresentamos dois exemplos ilustrativos de como um modelo LPV pode ser obtido através do procedimento de \emph{embedding}, além de aplicar estes conceitos ao estudo de caso.

\begin{remark}
Ressaltamos que o modelo LPV da Eq. \eqref{modeloLPV} não requer, por definição, nenhum tipo de dependência específica das matrizes $A(\rho), B(\rho), C(\rho)$ e $D(\rho)$  nos parâmetros $\rho$, ou seja, estas matrizes podem ser afins, polinomiais, fracionais, etc.
\end{remark}

\begin{remark}
Com intuito de simplificar as demonstrações futuras, consideramos, sem perda de generalidade, que o proxy LPV é apenas estado-dependente, ou seja, $\rho(k)\,=\,f_\rho(x(k))$, tal como no estudo de caso do \textit{gas-lift}. Ademais, negligenciaremos, doravante, as variáveis de saída, concentrando-nos apenas nos sinais de controle e de estados. Na Seção \ref{sec6}, retomamos a variável $y(k)$ para discutirmos o tema de rastreamento de referências.
\end{remark}

\begin{example}
Consideremos o caso de amortecedores semi-ativos (eletro-reológicos), tal como descritos em \citep{menezes2020development}. Esses sistemas apresentam um comportamento inerentemente histerético, representado por uma função tangencial hiperbólica. Assim, as dinâmicas da variação da distensão da suspensão ($z_d$) é dada, simplificadamente, por: $\dot{z}_d(k+1) = \dot{z}_d(k) + m\tanh \left(a_1\dot{z}_d(k) + a_2\dot{z}_d(k-1)\right)u(k)$, sendo $m,a_1$ e $a_2$ parâmetros fixos e $u$ um sinal relacionado a um campo elétrico controlável. Note, neste caso que, uma vez que $\dot{z}_d(k)$ é medido para toda amostra $k\,\geq\,0$, pode-se compor um modelo LPV do tipo $\dot{z}_d(k+1) = \dot{z}_d(k) + m\rho(k)u(k)$, no qual $\rho(k) \,=\, f_\rho(\dot{z}_d(k)) \,= \,\tanh\left(a_1\dot{z}_d(k) + a_2\dot{z}_d(k-1)\right)$.  
\end{example}

\begin{example}
Consideremos as dinâmicas do nível $h(k)$ de um tanque cilíndrico com um furo redondo aberto em sua base e com uma vazão de entrada $u$, descritas pela seguinte relação $h(k+1) = h(k) + a_1u(k) - a_2\sqrt{2gh(k)}$, sendo $a_1$ e $a_2$ parâmetros fixos. Neste caso, uma vez que $h(k)$ é mensurável, podemos usar $\rho(k) \,= \,f_\rho(h(k)) \,=\,\frac{\sqrt{2gh(k)}}{h(k)}$; note que o nível de água no tanque é considerado implicitamente, aqui, como sempre positivo. Assim, obtemos o seguinte modelo LPV: $h(k+1) = h(k) + a_1u(k) - a_2\rho(k)h(k)$.
\end{example}

\begin{tutorialstep}
Retomemos o processo de \textit{gas-lift}. Este sistema opera sob um período de amostragem de $T_s \,= \,5\, \text{s}$. Assim, podemos obter um modelo LPV através de uma inclusão diferencial exata do modelo não linear obtido através da discretização Euler da Eq. \eqref{EDOsMassa}. Usando as variáveis de agendamento apresentadas a seguir, obtemos um modelo LPV discreto tal como na Eq. \eqref{modeloLPV}, com matrizes postas nas Eqs. \eqref{MatrizAeC}-\eqref{MatrizBeD}:
\begin{eqnarray} 
\label{ValordosRho}
\left\{\begin{array}{ccl}
\rho_1(k)&=&\sqrt{\max(0,P_{fg}-\left(T_aR(V_aM_g)^{-1}+gH_aV_a^{-1}\right) x_1(k))}\, \text{,}\\ 
\rho_2(k)&=&\sqrt{x_1(k)}\, \text{,}\\
\rho_3(k)&=&\sqrt{\max(0,\left(\frac{T_aR}{V_aM_g}+\frac{gH_a}{V_a}\right)x_1(k)-(\frac{T_tR\mu_o}{M_g}\frac{x_2(k)}{V_t\mu_o-x_3(k))}+\frac{g}{A_t}(x_2(k)+x_3(k)))}\frac{1}{x_3(k)}\, \text{,}\\ 
\rho_4(k)&=&\max(0,P_r-(\frac{T_tR\mu_o}{M_g}\frac{x_2(k)}{V_t\mu_o-x_3(k)}+\frac{g}{A_t}(x_2(k)+x_3(k))+\mu_ogH_{fp}))\frac{1}{x_2(k)}\, \text{,}\\ 
\rho_5(k)&=&x_2(k)\left(\sqrt{x_2(k)+x_3(k)}\right)^{-1}\, \text{,}\\ 
\rho_6(k)&=&x_3(k)\left(\sqrt{x_2(k)+x_3(k)}\right)^{-1}\, \text{,}\\ 
\rho_7(k)&=&\sqrt{\max\left(0,T_tR\mu_oM_g^{-1}x_2(k)\left(V_t\mu_o-x_3(k)\right)^{-1}-P_s\right)}\, \text{,}\\ 
\rho_8(k)&=&\left(V_t\mu_o-x_3(k)\right)^{-1}\, \text{.} \end{array}\right.
\end{eqnarray}
\end{tutorialstep}

\begin{eqnarray}
\label{MatrizAeC}
A(\rho(k))&=&\begin{bmatrix}
1 & 0 & -T_sK_{inj}\sqrt{\frac{1}{V_a}(1+\frac{gH_aM_g}{T_aR})}\rho_2(k)\rho_3(k)\\ 
0 & 1+T_sg_{go}i_P\rho_4(k) & T_sK_{inj}\sqrt{\frac{1}{V_a}(1+\frac{gH_aM_g}{T_aR})}\rho_2(k)\rho_3(k)\\ 
0 & T_si_P\rho_4(k) & 1
\end{bmatrix} \, \text{,} \\
B(\rho(k))&=&\begin{bmatrix}
T_sK_{cp}\sqrt{\frac{1}{V_a}(1+\frac{gH_aM_g}{T_aR})}\rho_1(k)\rho_2(k) & 0\\ 
0 &-T_s\frac{K_{cgl}}{\sqrt{V_t}}\rho_5(k)\rho_7(k)\\ 
0 &-T_s\frac{K_{cgl}}{\sqrt{V_t}}\rho_6(k)\rho_7(k)
\end{bmatrix} \; \text{,}\; \\
C(\rho(k))&=&\begin{bmatrix}
0 & \frac{g}{A_t}+\frac{T_tR\mu_o}{M_g}\rho_8(k)& \frac{g}{A_t}\\ 
0 & 0 & 0
\end{bmatrix} \; \text{,}\\ 
\label{MatrizBeD}
D(\rho(k))&=&\begin{bmatrix}
0 & 0\\ 
K_{cp}\sqrt{\frac{1}{V_a}(1+\frac{gH_aM_g}{T_aR})}\rho_1(k)\rho_2(k) & 0
\end{bmatrix}\;\text{.}
\end{eqnarray}

\section{Estimando parâmetros futuros}
\label{sec3}

\noindent Uma vez obtida a representação LPV do processo controlado, seguindo as etapas descritas na Seção \ref{sec2}, passamos a utilizá-la para a \emph{predição} das dinâmicas futuras do sistema, ao longo de um horizonte futuro de $N_p$ passos, para fins do projeto de um algoritmo MPC. Para tal predição, calculamos os estados futuros $x(k+j), \, \forall \, j \in \, \mathbb{N}_{[1,N_p]}$ que dependem, dada a estrutura LPV da Eq. \eqref{modeloLPV}, dos valores futuros dos parâmetros de agendamento $\rho(k+j-1), \, \forall \, j \in \, \mathbb{N}_{[1,N_p]}$. Entretanto, apenas o sinal de agendamento amostrado $\rho(k)$ é conhecido a cada instante de tempo.

Neste caso, a \emph{segunda etapa} do projeto consiste em desenvolver uma lei (de baixo custo numérico) para extrapolar/estimar as trajetórias futuras dos parâmetros de agendamento. Na literatura, podemos encontrar três principais alternativas para esse procedimento: 
\begin{itemize}
    \item  O uso de "congelamento", ou seja, considerar que os parâmetros de agendamento não se alterarão ao longo do horizonte de predição, do ponto de vista de cada instante de amostragem: $\rho(k+j|k) \,=\, \rho(k), \forall j \, \in \, \mathbb{N}_{[1,N_p-1]}$. Essa abordagem, apesar de simplória e, de certa forma, ingênua, pode resultar em desempenhos práticos de relevância,  c.f. \citep{morato2018design,alcala2020autonomous}. Todavia, devido à grande incerteza de predição suscitada, a abordagem resulta no projeto de algoritmos MPC mais conservadores\footnote{Devido à incerteza incluída implicitamente ao modelo de predição por congelamento, o projeto dos algoritmos MPC, neste caso, deve levar em conta a estabilização para todos os valores possíveis  dos sinais de agendamento futuros, através de argumentos de robustez.}; 
    \item O uso de estimativas com base na iteração do problema de otimização MPC de forma sequencial, tal como originalmente proposto por \citet{cisneros2017fast} e amplamente utilizado desde então, c.f. \citep{cisneros2019wide}. A ideia, nesse caso, é usar o proxy $\rho(k+j|k) \,=\, f_\rho(x(k+j|k))$ recursivamente, no qual a sequência de estados futuros $x(k+j|k), \, \forall j \, \in \, \mathbb{N}_{[1,N_p-1]}$ é obtida da ultima iteração do problema de otimização. Tal como elaborado em \citep{cisneros2019wide}, pode-se demonstrar que a trajetória predita de parâmetros futuros converge para a trajetória real, ao passo que o programa de otimização torna-se mais complexo (resolve-se $n_i$ QPs por amostra);
    \item O uso de extrapolação com base em uma aproximação da função $f_\rho(\cdot)$, tal como proposto recentemente em \citep{morato2022sufficient}. A fim de suavizar o cômputo numérico da abordagem de \citet{cisneros2017fast}, esse método também garante a convergência da trajetória predita de parâmetros futuros para a trajetória real, ao passo que baseia-se apenas em operadores lineares. Neste artigo, usaremos essa última metodologia.
\end{itemize}

Doravante, consideramos $\hat{P}_k$ como a "trajetória futura (predita) de agendamento", ou seja, col$\{\rho(k|k)\,,\,\rho(k+1|k)\,,\, \dots \, , \, \rho(k+N_p-1|k)\}$. Similarmente, tomamos $\hat{X}_k$ como col$\{x(k+1|k)\,\dots\, x(k+N_p|k)\}$ como os estados futuros (preditos). Assim, prosseguimos detalhando a ideia principal disposta em \citep{morato2022sufficient}: construir $\hat{P}_k$ recursivamente, com base em $\hat{P}_{k-1}$, e aproveitando de uma expansão de Taylor de primeira ordem da função $\rho(k+j|k)\,=\,f_\rho(x(k+j|k)$. Ressaltamos que a principal vantagem deste método é que, além de produzir estimativas \emph{convergentes}, com erro residual limitado, apenas operadores lineares são necessários.

O método constrói-se da seguinte forma\footnote{Aqui, consideramos $\delta x (k) \,= \, x(k+1)-x(k)$ e, portanto, $\delta\hat{X}_k \,:=\, [\delta x(k|k) \, \dots \, \delta x(k+N_p-1|k)]^T$.}: consideramos, primeiramente, a seguinte expansão de Taylor:
\begin{eqnarray}\label{TaylorEqExp}
    f_\rho(x) &=& f_\rho(\breve{x}) + f_\rho^\partial (x-\breve{x}) +\xi_\rho \, \text{,}
\end{eqnarray}
\noindent sendo $\breve{x}\,\in\,\mathcal{X}$ um ponto de expansão admissível, $f_\rho^\partial \,:=\,\left.\frac{\partial f_\rho(x)}{\partial x}\right|_{x=\breve{x}}$ a derivada do proxy LPV, considerando este de classe $\mathcal{C}^1$, e $\xi_\rho$ um termo residual que incorpora o erro de aproximação. Assim, desconsiderando os termos residuais, podemos re-escrever toda a trajetória futura de agendamento, com base na Eq. \eqref{TaylorEqExp}, da seguinte forma:
\begin{eqnarray} \label{rhoshorizon}
\rho(k+1|k) &=& \rho(k|k-1) + f_\rho^\partial (k)\delta x (k|k) \,
              \text{,} \\ \nonumber
\rho(k+2|k) &=& \rho(k+1|k-1) + f_\rho^\partial (k+1)\delta x (k+1|k) \,
              \text{,} \\ \nonumber
&\vdots& \\ \nonumber
\rho(k+N_p-1|k) &=& \rho(k+N_p-2|k-1) \\ \nonumber &+& f_\rho^\partial (k+N_p-2)\delta
                                             x (k+N_p-2|k) \, \text{.}
\end{eqnarray}

Na Eq. \eqref{rhoshorizon}, os termos $f_\rho^\partial(k+j)$ e $\delta x (k+j|k)$, para todo $j\,\in\,\mathbb{N}_{[1,N_p-2]}$, são desconhecidos, Todavia, tal como recomenda-se em \citep{morato2022sufficient}, essas variáveis são substituídas, respectivamente, por $f_\rho^\partial(k)$ e $\delta x (k+j|k-1)$, ou seja: (a) considera-se, implicitamente, que a derivada parcial $f_\rho^\partial(k+j)\,:=\,\left.\frac{\partial f_\rho (x)}{\partial x}\right|_{x(k+j)}$ é constante e igual a $f_\rho^\partial(k)$, de valor conhecido no instante amostrado $k$; e (b) a trajetória futura predita para as variações dos estados, ao longo do horizonte de predição, é recuperada da última iteração do programa de otimização, do instante $k-1$.

Através desse argumento, podemos estabelecer a seguinte evolução para os parâmetros de agendamento: $\rho(k+j|k) \approx \rho(k+j-1|k-1) + f_\rho^\partial(k) \delta x(k+j-1|k-1), \, \forall j \, \in \, \mathbb{N}_{[1,N_p-1]}$. Em forma vetorial, a trajetória futura (predita) de agendamento $\hat{P}_k$ pode ser obtida com base na trajetória predita no instante anterior, ou seja $\hat{P}_{k-1}$, somada a um termo de correção $f_\rho^\partial(k)\delta \hat{X}_k$, o que resulta no seguinte procedimento de extrapolação:
\begin{eqnarray}
\label{RecursiveProposed}
\hat{P}_k &=& \hat{P}_{k-1} + f_\rho^\partial(k)\delta \hat{X}_k\, \text{,}
\end{eqnarray}
\noindent considerando a apropriação dada por: $\delta \hat{X}_k \,\approx\,\text{col}\{\delta x(k|k)\,,$ $\delta x(k+1|k-1)\,,\,\dots\, , \, \delta x (k+N_p-2|k-1)\}$.

Em termos de implementação da Eq. \eqref{RecursiveProposed}, ressaltamos os seguintes detalhes:
\begin{itemize}
    \item No instante de amostra inicial, ou seja $k\,=\,0$, não há estimativa prévia disponível. Portanto, toma-se, simplesmente, a estimativa inicial como $\hat{P}_0 = \left[\begin{array}{cccc}\rho^T(0) & \rho^T(0) & \dots & \rho^T(0)\end{array}\right]^T$, ou seja um vetor com $N_p$ entradas repetidas do parâmetro de agendamento medido $\rho(0)$. 
    \item Para as amostras seguintes:
    \begin{enumerate}
        \item O vetor de variações dos estados $\delta \hat{X}_k$ é construído com base na solução do algoritmo de otimização da amostra anterior, $\delta \hat{X}_{k-1}$;
        \item A derivada parcial é avaliada para o instante amostral atual, ou seja, $f_\rho^\partial(k)\,=\,\left.\frac{\partial f_\rho(x)}{\partial x}\right|_{x(k)}$;
        \item Por fim, a trajetória futura de agendamento predita é calculada através da Eq. \eqref{RecursiveProposed}.
    \end{enumerate}
\end{itemize}

Enfatizamos que na Eq. \eqref{RecursiveProposed}, o termo residual $\xi_\rho$, proveniente da expansão de Taylor da Eq. \eqref{TaylorEqExp} foi negligenciado. Todavia, resultados concretos estão disponíveis na literatura demonstrando como estes termos tendem a zero, com o passar das amostras, c.f. \citep{morato2021fast,morato2022LPVSIO, morato2022sufficient}. Nestas referências, demonstra-se como obter um limite superior para a norma do termo residual, como também condições de suficiência para convergência do método de extrapolação.

\begin{remark}
A grande vantagem do método desenvolvido nestes artigos consiste no fato desse gerar estimativas convergentes, com erros residuais limitados, com base, unicamente, em operadores \emph{lineares}. Desta forma, a trajetória predita $\hat{P}_k$ possui baixo custo numérico e resulta em um erro de predição reduzido, permitindo a síntese de controladores preditivos rápidos e pouco conservadores.
\end{remark}

\begin{tutorialstep}
    Em relação ao estudo de caso do processo de \textit{gas-lift}, mencionamos a seguir alguns detalhes quanto à implementação da estratégia de estimativa dos parâmetros LPV futuros. Para tal, tomamos como exemplo ilustrativo o quinto sinal de agendamento, $\rho_5(k) \, = \, f_{\rho_5}(x(k))$ na Eq. \eqref{ValordosRho}. Ressaltamos que a derivada total deste sinal, em relação aos estados, é dada por:
    \begin{eqnarray}
        f_{\rho_5}^\partial \, :=\, \frac{\partial f_{\rho_5}(x)}{\partial x} &=& \left[\begin{array}{ccc} 0 &  \left(\frac{1}{\sqrt{(x_2 + x_3)}} - \frac{x_2}{2(x_2 + x_3)^{\frac{3}{2}}}\right) & -\frac{x_2}{2(x_2 + x_3)^{\frac{3}{2}}}]
        \end{array}\right] \, \text{.}
    \end{eqnarray}
    Portanto, a lei de recursão resultante é dada por:
    \begin{eqnarray}
       \left\{\begin{array}{ccl}
            \rho_5(k+1|k)&=& \rho_5(k|k-1) + f_{\rho_5}^\partial(k)\delta x (k|k-1) \, \text{,}\\
            &\vdots& \\ 
            \rho_5(k+N_p-1|k)&=& \rho_5(k+N_p-2|k-1) + f_{\rho_5}^\partial\delta x (k+N_p-2|k-1)\, \text{.}
       \end{array}\right.
    \end{eqnarray}
\end{tutorialstep}

\section{MPC LPV: Aspectos de implementação}
\label{sec4}

Com base na predição da trajetória futura de agendamento $\hat{P}_k$, podemos prosseguir com detalhes da síntese do controlador preditivo em si. Portanto, nessa Seção, discutimos aspectos de projeto e de implementação. 

Retomamos o problema: buscamos um algoritmo controle MPC para controlar o sistema não linear, originalmente descrito pela Eq. \eqref{sisnonlin}, com base em sua representação LPV apresentada nas Eqs. \eqref{modeloLPV}-\eqref{modeloLPV2}. Para tal, consideramos as restrições operacionais de admissibilidade da Eq. \eqref{XUYcalconstraints}, que devem ser respeitadas para todo instante de tempo. Ademais, levando em conta a metodologia do controle preditivo, preocupamo-nos com as trajetórias futuras das variáveis de interesse ao longo de um horizonte futuro de $N_p$ amostras. Nesse sentido, consideramos a trajetória de agendamento futura $\hat{P}_k$.

Relembramos que, no contexto de MPC, uma nova lei de controle ótima $u(k)$ é gerada a cada amostra discreta de tempo $k$, através da resolução de um problema de otimização, denominado doravante de $\mathfrak{O}_k$. Portanto, o primeiro aspecto de interesse para a síntese de tal controlador preditivo é a lei de predição a ser usada em cada $\mathfrak{O}_k$, ou seja, qual é a relação entre as variáveis futuras e as variáveis medidas na amostra atual. 

Assim, além do vetor de estados futuros (preditos) $\hat{X}_k$ e da trajetória (predita) de agendamento $\hat{P}_k$, consideramos: a sequência\footnote{Negligenciamos o uso da notação com circunflexo no vetor $U_k$, uma vez que esse vetor não é apenas um vetor de variáveis preditas, como também a variável de decisão do problema de otimização $\mathfrak{O}_k$.} de sinais de controle a ser aplicada, dada por $U_k \,:=\, \text{col}\{u(k|k) \, , \, \dots \, , \, u(k+N_p-1|k)\}$, como o vetor das saídas futuras (preditas), dado por $\hat{Y}_k \,:=\, \text{col}\{y(k|k) \, , \, \dots \, , \, y(k+N_p-1|k)\}$. Com base nessa construção vetorial, podemos descrever as variáveis futuras através da seguinte lei de predição, obtida pela expansão da Eq. \eqref{modeloLPV} ao longo do horizonte:
\begin{eqnarray}\label{XkdeviationEq}
\left\{\begin{array}{ccl} X_k &=& \mathcal{A}(\hat{P}_{k})x(k) + \mathcal{B}(\hat{P}_k)U_k \, \text{,} \\ Y_k &=& 
\mathcal{C}(\hat{P}_{k})x(k) + \mathcal{D}(\hat{P}_k)U_k \, \text{.}\end{array}\right.
\end{eqnarray} 
\noindent As matrizes de predição $\mathcal{A}(P_k) \in \mathbb{R}^{(n_x N_p) \times n_x}$, $\mathcal{B}(P_k) \in \mathbb{R}^{(n_x N_p) \times n_u}$, $\mathcal{C}(P_k) \in \mathbb{R}^{(n_y N_p) \times n_x}$, e $\mathcal{D}(P_k) \in \mathbb{R}^{(n_y N_p) \times n_u}$, apresentadas, respectivamente, nas Eqs. \eqref{PredicAPk}-\eqref{PredicDPk}, mantêm estrutura e forma a cada amostra discreta $k$ e, portanto, podem ser calculadas de forma eficiente, em termos computacionais.

Com base na lei de predição colocada na Eq. \eqref{XkdeviationEq}, a implementação de um algoritmo de controle preditivo LPV, portanto, se dá através da resolução do seguinte problema de otimização $\mathfrak{O}_k$, a cada amostra de tempo discreta $k\,\geq\,0$:
\begin{eqnarray}
\label{TheMPC}
    \min_{U_k} & J(X_k,Y_k,U_k)\, \text{,} \\ \nonumber
\text{t.q.}: &\left\{ \begin{array}{rcl}
    		X_k &=& \mathcal{A}(\hat{P}_k)x(k) +  \mathcal{B}(\hat{P}_k)U_k \, \text{,} \\
        Y_k &=& \mathcal{C}(\hat{P}_k)x(k) +  \mathcal{D}(\hat{P}_k)U_k \, \text{,} \\
        x(k+j|k) &\in& \mathcal{X}, \, j \in \mathbb{N}_{[1,N_p]}, \\
    	y(k+j-1|k) &\in& \mathcal{Y}, \, j \in \mathbb{N}_{[1,N_p]}, \\
        u(k+j-1|k) &\in& \mathcal{U}, \, j \in \mathbb{N}_{[1,N_p]}, \\ x(k+N_p|k) &\in& \mathbf{X}_f.
    \end{array}
    \right. 
\end{eqnarray}

O problema de otimização $\mathfrak{O}_k$, tal como colocado na Eq. \eqref{TheMPC}, contém as restrições de admissibilidade do processo (os sinais de estado, saída e controle considerados devem respeitar as restrições físicas do sistema), tal como uma \emph{restrição terminal} para a trajetória dos estados ($x(k+N_p|k)$ deve pertencer ao \emph{conjunto terminal} $\mathbf{X}_f$). Consideramos, nesse problema de otimização, a função de custo quadrática dada por:
\begin{eqnarray}
    \label{TheCustoJk}
    J_k &=& J(X_k,U_k) \,:=\, V\left(x(k+N_p|k)\right)\\ \nonumber &+& \sum_{j=1}^{N_p-1}\ell\left(x(k+j|k), u(k+j-1|k)\right) \text{,}
\end{eqnarray}
\noindent sendo $V(\cdot)$ um \emph{custo terminal} quadrático e $\ell(x,u)$ um \emph{custo de etapa} quadrático, genericamente ilustrado por:
\begin{eqnarray}\label{CustoDeEtapa}
    \ell(x,u) &:=& \Vert x \Vert_Q^2  +\Vert u \Vert_R^2 \, \text{,}
\end{eqnarray}
\noindent sendo $Q$ e $R$ matrizes de ponderação definidas positivas, usadas para impor o objetivo de controle ao processo.

Assumindo que as restrições de admissibilidade são convexas, e uma vez que $\mathfrak{O}_k$ possui um custo quadrático $J_k$, para toda amostra discreta $k\,\geq\,0$, tal como uma lei de predição linear (note que Eq. \eqref{XkdeviationEq} é linear para cada vetor $\hat{P}_k$), torna-se fato que $\mathfrak{O}_k$ é um \textbf{programa de otimização quadrático} (QP), podendo ser resolvido em \emph{tempo real}, na escala dos microssegundos\footnote{O tempo de resolução também depende do processador e do \textit{solver}.}. 

Ressaltamos que a solução amostrada desse problema, dita $U_k^\star$, é uma sequência de ações de controle, cuja primeira entrada, $u^\star(k|k)$, é aplicada ao processo. A implementação, de fato, se dá na seguinte ordem:
\begin{itemize}
    \item Etapa de projeto (\textit{offline}):
    \begin{enumerate}
        \item Ajusta-se o custo de etapa $\ell(x,u)$ conforme o objetivo de controle, ajustando as matrizes de ponderação $Q$ e $R$;
        \item Calculam-se os ditos \emph{ingredientes terminais}: o custo terminal $V(x(k+N_p|k))$ e o conjunto terminal $\mathbf{X}_f$ (detalhes na Seção \ref{sec5}).
    \end{enumerate}
    \item Durante a implementação (\textit{online}), para todo $k\,\geq\,0:$
      \begin{enumerate}
        \item Mede-se o estado atual $x(k)$ e calcula-se o parâmetro de agendamento correspondente $\rho(k)$, através do proxy de agendamento $f_\rho(\cdot)$;
        \item Obtém-se a trajetória futura de agendamento $\hat{P}_k$ através da lei de extrapolação da Eq. \eqref{RecursiveProposed} (etapas apresentadas no final da Seção \ref{sec3});
        \item Monta-se a lei de predição conforme a Eq. \eqref{XkdeviationEq};
        \item Resolve-se o programa de otimização quadrático $\mathfrak{O}_k$, apresentado na Eq. \eqref{TheMPC}, cuja solução é $U_k^\star$;
        \item Aplica-se a lei de controle preditiva LPV $u^\star(k|k)$.
    \end{enumerate}
\end{itemize}

\begin{eqnarray}
\label{PredicAPk}
\mathcal{A}(\hat{P}_k) &:=&\left[\begin{array}{c} A(\rho(k|k)) \\ A(\rho(k+1|k))A(\rho(k|k)) \\ \vdots \\ A(\rho(k+N_p-1|k))A(\rho(k+N_p-2|k))\dots A(\rho(k|k))\end{array}\right]
  \, \text{,} \\
  \label{PredicBPk}
\mathcal{B}(\hat{P}_k) &:=&\left[\begin{array}{ccc} B(\rho(k|k)) & 0 & \dots \\ A(\rho(k+1|k))B(\rho(k|k)) &  B(\rho(k+1|k)) & \dots \\ &\ddots &\\ A(\rho(k+N_p-1|k))\dots A(\rho(k+1|k))B(\rho(k|k)) & A(\rho(k+N_p-1|k))\dots A(\rho(k+2|k))B(\rho(k+1|k))& \dots\end{array}\right]
  \, \text{,} \\
  \label{PredicCPk}
\mathcal{C}(\hat{P}_k) &:=&\left[\begin{array}{c} C(\rho(k|k)) \\ C(\rho(k+1|k))A(\rho(k|k)) \\ \vdots \\ C(\rho(k+N_p-1|k))A(\rho(k+N_p-2|k))\dots A(\rho(k|k))\end{array}\right]
  \, \text{,} \\
  \label{PredicDPk}
\mathcal{D}(\hat{P}_k) &:=&\left[\begin{array}{ccc} D(\rho(k|k)) & 0 & \dots \\ C(\rho(k+1|k))B(\rho(k|k)) &  D(\rho(k+1|k)) & \dots \\ &\ddots &\\ C(\rho(k+N_p-1|k))\dots C(\rho(k+1|k))B(\rho(k|k)) & C(\rho(k+N_p-1|k))\dots C(\rho(k+2|k))B(\rho(k+1|k))& \dots\end{array}\right]
  \, \text{.} 
\end{eqnarray}

\section{Ingredientes terminais e certificados}
\label{sec5}

Os ingredientes terminais $V(\cdot)$ e $\mathbf{X}_f$ são utilizados amplamente na literatura de controle preditivo, c.f. \citep{limon2018nonlinear,morato2023IQCstabilizing}, visando garantir estabilidade em malha fechada e uma otimização com \emph{factibilidade recursiva}\footnote{Esta propriedade dita que uma vez que o problema de otimização inicial $\mathfrak{O}_0$ é \emph{factível}, ou seja possuí, de fato, uma solução $U_0^\star$, toda otimização futura $\mathfrak{O}_k$, para todo $k\,>\,0$, também será factível. Usa-se, costumeiramente, a solução $U^\star_{k}$ como candidata para a construção da solução $U^\star_{k+1}$.}. Nesta Seção, portanto, detalhamos as etapas necessária para síntese desses elementos. 

\begin{remark}
Ressaltamos que, visando simplicidade nas discussões seguintes, desconsideramos a presença de perturbações não nulas ou erros de predição. Discussões sobre o caso robusto completo podem ser consultadas em \citep{cunha2022robust}. Ademais, consideramos o problema de regulação, ou seja, o objetivo de controle diz respeito à estabilização da trajetórias dos estados à origem. O caso de rastreamento de referências é debatido na Seção \ref{sec6}.
\end{remark}

\begin{assumption} 
\label{hipo1} 
O custo de etapa é uma função uniformemente contínua e definida positiva de tal forma que $\ell(x,u) \,\geq\, \alpha_\ell(\Vert x \Vert)$ e $|\ell(x_1,u_1)-\ell(x_2,u_2)| \,\leq\, \lambda_e(\Vert x_1-x_2\Vert) +\lambda_u(\Vert u_1-u_2\Vert)$, sendo $\alpha_\ell$, $\lambda_x$ e $\lambda_u$ funções de classe $\mathcal{K}$; implicitamente, define-se que $\ell(0,0)=0$.
\end{assumption}

\begin{assumption}
\label{hipo2}
O sistema não linear, representado pelo modelo LPV da Eq. \eqref{modeloLPV}, é estabilizável e, portanto, existe uma realimentação de estados estabilizante e admissível, ou seja, um ganho $\kappa_t$ de tal forma que $u(k) = \kappa_t x(k) \, \in \, \mathcal{U}\, ,\, \forall x(k) \, \in \, \mathbf{X}_t$, sendo esse um conjunto terminal positivo invariante para o sistema. Ademais, o custo terminal $V(x)$ é contínuo e positivo para todo $x\in\,\mathbf{X}_f$, de tal forma que existam as constantes $\beta\,>\,0$ e $\sigma\,>\,1$ impondo $V(x) \leq b|x|^\sigma, \forall x \,\in \, \mathbf{X}_f$, fato que implica na desigualdade $V(x(k+1))-V(x(k)\,\leq\, -\ell\left(x(k),\kappa_t x(k)\right)$, para todo $x(k) \in \mathbf{X}_f$.
\end{assumption}

Com base nas Hipóteses \ref{hipo1} e \ref{hipo2}, o Teorema \ref{Theo1}, adaptado de \citep{mayne2000constrained} para o caso LPV, apresenta as condições de suficiência para estabilidade em malha fechada e para que $\mathfrak{O}_k$ seja recursivamente factível. Ademais, o Teorema \ref{Theo2}, originalmente apresentado em \citep{morato2021robustzono}, oferece uma abordagem baseada em desigualdades matriciais lineares (LMIs) para o cômputo dos ingredientes terminais.

\begin{theorem}[Ingredientes terminais \citep{mayne2000constrained}.]\label{Theo1}
Supõe-se que as Hipóteses \ref{hipo1} e \ref{hipo2} são válidas e que a lei de controle preditiva LPV $u^\star(k)$ é dada pela resolução do problema de otimização quadrático $\mathfrak{O}_k$ tal qual apresentado na Eq. \eqref{TheMPC}, com função de custo $J_k$ dado pela Eq. \eqref{TheCustoJk}, assumindo um custo de etapa quadrático tal qual apresentado na Eq. \eqref{CustoDeEtapa}. Então, estabilidade assinótica da trajetória dos estados, em direção à origem, é estabelecida desde que as seguintes condições são válidas para toda variável de agendamento $\rho \, \in \, \mathcal{P}$:
\begin{enumerate}[label={[C\arabic*]}]
    \item A origem $x\,=\,0$ está contida no interior do conjunto terminal $\mathbf{X}_f$;
    \item $\mathbf{X}_f$ é um conjunto invariante positivo para o modelo LPV dado pelas Eqs. \eqref{modeloLPV}-\eqref{modeloLPV2}, sob ação da lei de controle admissível $u(k)\,=\,\kappa_t x(k)$;
    \item A desigualdade de Lyapunov é verificada para os estados contidos dentro do conjunto invariante terminal, ou seja, para todo $x(k) \, \in \, \mathbf{X}_f$ e todo parâmetro de agendamento $\rho\, \in \, \mathcal{P}$, é verificado que:
    \begin{eqnarray}
        \nonumber
        V\left(x(k+1)\right) - V\left(x(k)\right) &\leq& -\ell\left(x(k),\kappa_t x(k)\right) \, \text{.}
    \end{eqnarray}
    \item A imagem do controle terminal é admissível para todo estado dentro do conjunto invariante, ou seja, para todo $x(k) \, \in \, \mathbf{X}_f$, verifica-se que $u(k) \, \in \, \mathcal{U}$;
    \item O conjunto terminal $\mathbf{X}_f$ é um subconjunto admissível, ou seja, $\mathbf{X}_f\,\subseteq \, \mathcal{X}$.
\end{enumerate}
Ademais, sendo a solução inicial do problema de otimização $U_0^\star$ admissível, o algoritmo MPC LPV da Eq. \eqref{TheMPC} é recursivamente factível.
\end{theorem}

Para satisfazermos as condições colocadas no Teorema \ref{Theo1}, consideramos um custo terminal quadrático, ou seja $V(x) \,:=\, \Vert x \Vert_P^2$, sendo $P\,=\,P^T$ uma matriz de ponderação definida positiva. Ademais, tomamos o conjunto terminal com o seguinte sub-conjunto de nível do custo terminal: $\mathbf{X}_f \,:=\, \{x \in \mathbb{R}^{n_x} \, | \,  \Vert x \Vert_P^2 \leq 1 \}$.

\begin{theorem}[Projeto com LMIs \citep{morato2021robustzono}]\label{Theo2} $\text{}$\\
 Supõe-se que as Hipóteses \ref{hipo1} e \ref{hipo2} são válidas. Então, as condições de suficiência (C1)-(C5) apresentadas no Teorema \ref{Theo1} são satisfeitas desde que exista uma matriz simétrica definida positiva $P \in \mathbb{R}^{n_x \times n_x}$ e uma matriz retangular $W \in \mathbb{R}^{n_u \times n_x}$ de tal forma que $Y \,:=\, P^{-1} \,>\, 0$, $W \,=\, \kappa_t Y$ e que as LMIs \eqref{LMI1}-\eqref{LMI3} são verificadas sob a minimização do argumento $\log \text{det}\left(Y\right)$ para todo parâmetro de agendamento 
$\rho \in \mathcal{P}$.
\begin{eqnarray}
\label{LMI1}
\left[\begin{array}{cc|cc} Y &  \star & \star & \star\\
(A(\rho)Y + B(\rho)W) & Y &\star & \star \\
Y & 0 & Q^{-1} & \star \\  W & 0 & 0 & R^{-1}  \end{array}\right] \,\,\geq\,\, 0 &\text{,}& \\
\label{LMI2}
\left[\begin{array}{c|c} \overline{x}_j^2 & I_{\{j\},n_x}Y \\ \hline I_{\{j\}}Y' & Y \end{array} \right] \,\,\geq\,\, 0, \, j \, \in \, \mathbb{N}_{[1,n_x]} &\text{,}&\\
\label{LMI3}
\left[\begin{array}{c|c} \overline{u}^2_i & I_{\{i\},n_u}W
        \\ \hline \star & Y\end{array}\right] \,\,\geq\,\, 0  , \, i\, \in \, \mathbb{N}_{[1,n_u]}
                          \, &\text{.}& 
\end{eqnarray}
\end{theorem}

\begin{proofbreve}
Através de dois complementos de Schur, a LMI \eqref{LMI1} torna-se a desigualdade da condição (C3), suficiente, portanto, para (C2). Da forma elipsoidal de $\mathbf{X}_f$, verificamos (C1). Através de complementos de Schur aplicados nas LMIs \eqref{LMI2} e \eqref{LMI3}, verificamos (C5) e (C4).\hfill $\square$
\end{proofbreve}

\begin{remark}
Os ingredientes terminais calculados através das LMIs apresentadas no Teorema \ref{Theo2} garantem a factibilidade recursiva do algoritmo MPC LPV da Eq. \eqref{TheMPC}. Ademais, garante-se a estabilidade assintótica das trajetórias dos estados em direção à origem. Ressaltamos, todavia, que a LMI \eqref{LMI1} possui dimensão infinita, uma vez que deve ser verificada para todo $\forall \, \rho \, \in \, \mathcal{P}$. Na prática, podemos obter uma solução factível ao forçar a desigualdade matricial para um número suficiente de pontos dentro do conjunto $\mathcal{P}$. Depois, a solução obtida pode ser re-verificada para grade mais densa de pontos.
\end{remark}

\begin{proposition}[Factibilidade recursiva]
\label{proporecursive}
    Consideremos $Y$ como uma solução válida para o Teorema \ref{Theo2}. Então, dado qualquer estado $x(k) \in \mathbf{X}_f$, obtemos $x(k+1) = \left(A(\rho)+B(\rho)\kappa_t\right)x(k) \,\in \,\mathbf{X}_f$. Consideremos a sequência ótima inicial $U_0^\star \,= \,(u^\star(0|0),u^\star(1|0),\dots,u^\star(N_p-1|0))$, solução factível de $\mathfrak{O}_1$. Então, a sequência $U_1^c \,=\, (u^\star(1|0),\dots,u^\star(N_p-1|0),\kappa_t x(N_p|0))$ constitui uma solução candidata factível para o problema de otimização subsequente, $\mathfrak{O}_1$. Assim, o algoritmo MPC LPV torna-se recursivamente factível.
\end{proposition}

\begin{proposition}[Estabilidade assintótica]
\label{propostab}
   Consideremos $Y$ como uma solução válida para o Teorema \ref{Theo2}. Então, o sistema não linear descrito pelo modelo LPV da Eq. \eqref{modeloLPV}, em malha fechada com o algoritmo MPC operado através da Eq. \eqref{TheMPC}, apresenta uma trajetória de estados que converge assintoticamente em direção à origem. Assim, para qualquer condição inicial factível $x(0)$, fica imposto que $\Vert x(k)\Vert \leq \beta(\Vert x(0) \Vert,k)$, sendo $\beta$ uma função de classe a $\mathcal{K}$ que passa pela origem.
\end{proposition}

\section{Rastreamento de referências}
\label{sec6}

Ao longo deste trabalho, consideramos, por enquanto, o problema da regulação das trajetórias dos estados. Todavia, em muitos casos, é comum querermos impor como objetivo de controle o rastreamento de referências pelo sinal de saída. Para tal, consideramos algumas alterações ao algoritmo MPC LPV.

Continuaremos considerando que os estados $x(k)$ são mensuráeis para toda amostra de tempo discreta $k\,\geq\,0.$ Assim, podemos determinar pares $(x_r,u_r)$ de regime permanente que impõe a estabilização da saída em um dado ponto $y_r$. Na prática, podemos guiar as trajetórias dos estados até o regime permanente $x_r$ (invés da origem), fazendo com que a saída atinja o objetivo de rastreio $y_r$.

Tal como debatido em \citep{limon2018nonlinear,koehler2019nonlinear,morato2022predictive}, podemos considerar que existe uma combinação \emph{única} de estados e sinais de controle, para cada referência admissível $y_r\,\in\,\mathcal{Y}$, que garante que $\lim_{k\to+\infty} y(k) \, = \, y_{r}$. Dessa forma, para cada $y_r$ podemos obter um par entrada-estado $(x_r,u_r)$ que valide a seguinte condição necessária de igualdade:
\begin{equation}
\label{NPminxrur}
    \begin{bmatrix} x_r-(A(f_\rho(x_r))x_r - B(f_\rho(x_r))u_r) \\ C(f_\rho(x_r))x_r + D(f_\rho(x_r))u_r \end{bmatrix} = \begin{bmatrix}0_{n_x} \\ y_r\end{bmatrix}.
\end{equation}

\begin{remark}
    Em muitos casos, podemos obter funções analíticas explícitas que satisfazem a Eq. \eqref{NPminxrur}, do tipo $x_r\,=\,r_x(y_r)$ e $u_r\,=\,r_u(y_r)$, c.f. \citep{limon2018nonlinear}.  
\end{remark}

Por fim, levando em conta o conhecimento de cada par $(x_r,u_r)$ para cada referência admissível $y_r$, adaptamos o algoritmo MPC da seguinte forma:
\begin{itemize}
    \item Tomamos o custo de etapa para rasteiro referência como: $\ell(x,u) \, := \,\Vert x - x_r \Vert_Q^2 + \Vert u - u_r \Vert_R^2$. Ademais, o custo terminal é adaptado para $V(x)\,:=\, \Vert x - x_r \Vert_P^2$.
    \item O conjunto terminal invariante $\mathbf{X}_f$ é tomado em relação ao erro de rastreio $e(k)=x(k)-x_r$. Assim, utilizamos a restrição $\left(x(k+N_p|k) - x_r\right) \, \in \, \mathbf{X}_f$ no problema de otimização $\mathfrak{O}_k$ da Eq. \eqref{TheMPC}.
\end{itemize}

\section{Aplicação ao estudo de caso}
\label{sec7}

\noindent A seguir, exemplificamos a aplicação MPC com o ferramental LPV para o estudo de caso do processo de \textit{gas-lift}. Para fins de controle na operação de plataformas de petróleo, consideramos como variáveis de interesse a vazão de entrada de gás $w_{g,in}(k)$ com o objetivo de minimizar os custos associados ao uso de gás e a pressão no fundo do poço $P_{fp}(k)$ com o intuito de maximizar a produção de óleo. Os valores de referência de regime permanente, para essas variáveis, são ajustados de acordo com as necessidades específicas da operação de extração. Por exemplo, em um cenário em que se deseja apenas maximizar a produção de óleo, pode ser necessário injetar uma grande quantidade de gás para aumentar a diferença de pressão entre o reservatório e o tubo. No entanto, isso pode resultar em uma produção na saída do \textit{choke} predominantemente composta por gás, o que pode não ser desejável. Portanto, nas plataformas de petróleo, os valores de referência para essas variáveis são determinados por meio de um processo de otimização.

Especificamente, consideramos, neste estudo de caso, apenas o controle local de vazões e pressões, dados sinais de referências seguindo a metodologia descrita na Seção \ref{sec6}. Para o projeto, usamos: $N_p\,=\,6$,  $Q\,=\,\text{diag}(10^{-12}\;\;8\cdot10^{-3})$, $R\,=\,\text{diag}(1\;\;1)$, considerando $\ell(x,u)\,:=\, \Vert x-x_r \Vert_Q + \Vert u - u_r \Vert_R$.

\begin{figure}
\centering
\includegraphics[width=\linewidth]{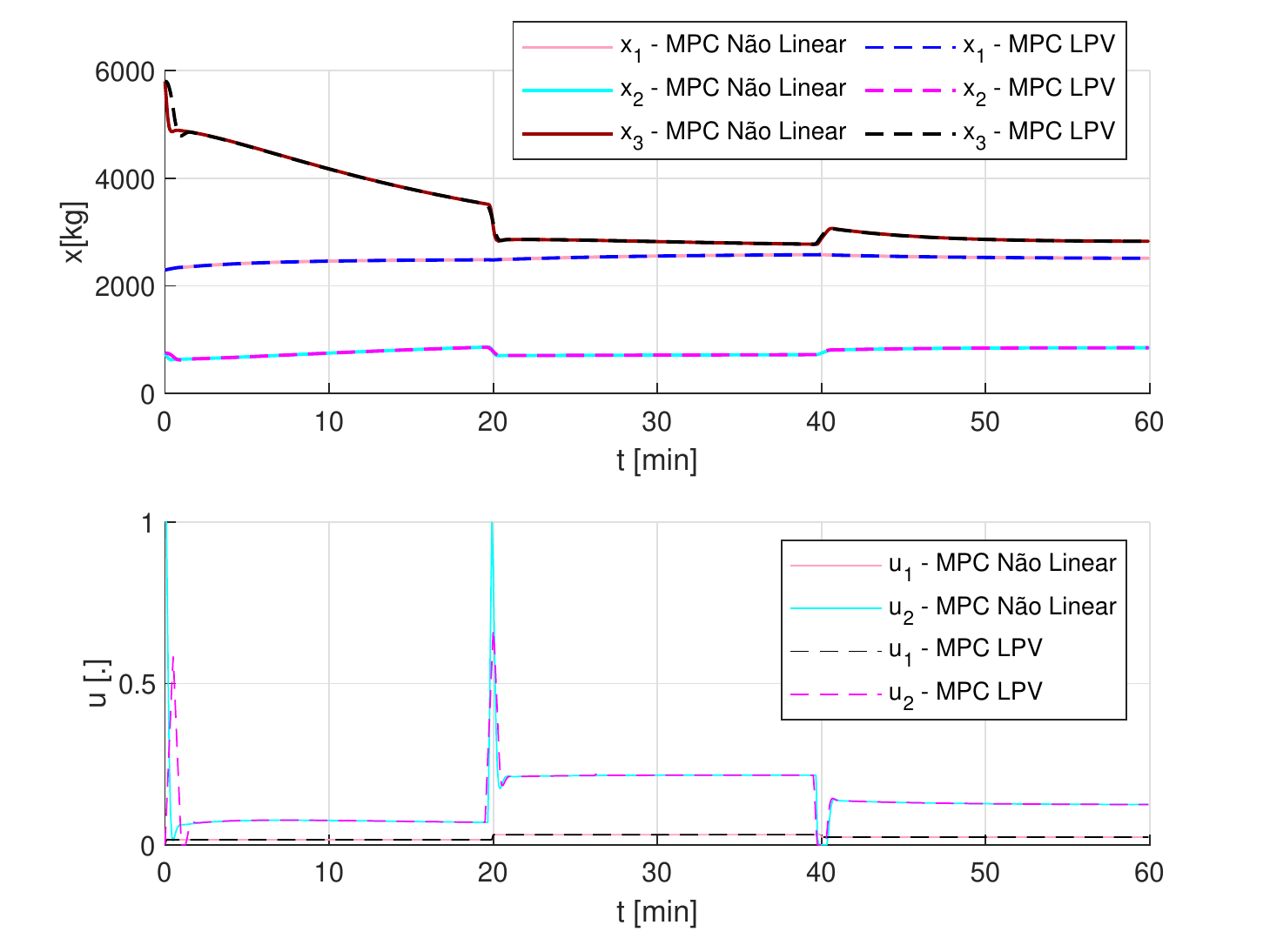}
\caption{Processo do \textit{gas-lift}: Estados e controle.}
\label{estados}
\end{figure}

\begin{figure}
\centering
\includegraphics[width=\linewidth]{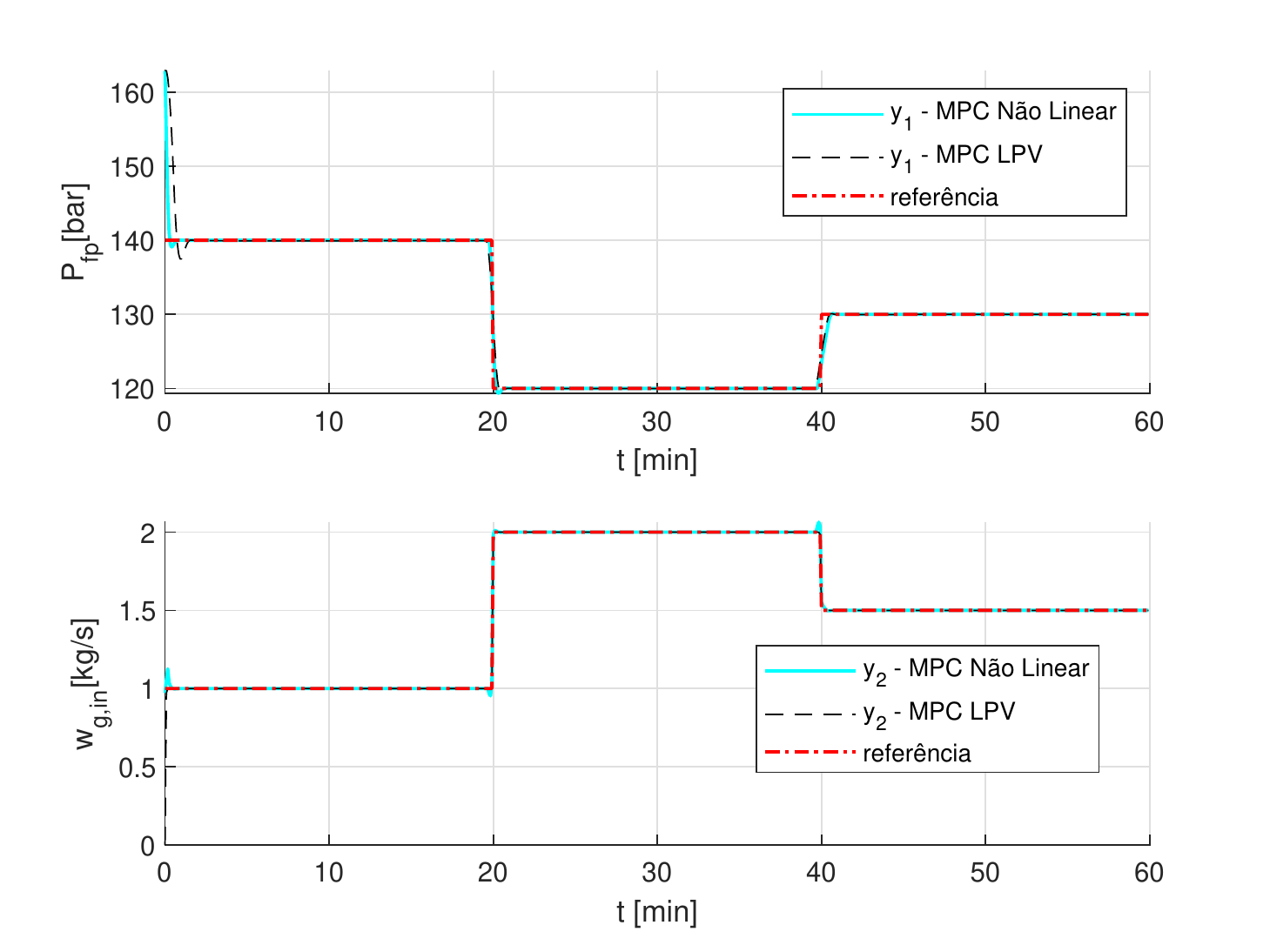}
\caption{Processo do \textit{gas-lift}: Saídas $P_{fp}$ e $w_{g,in}$.}
\label{saida1}
\end{figure}

Para ilustrarmos a eficácia da abordagem LPV, a comparamos
\footnote{Os resultados apresentados a seguir foram obtidos em simulação numérica operacionalizada com Matlab, Python, Yalmip e \textit{Gurobi} em um PC de 2.1 GHz e 8 GB RAM.}. com um software NMPC de ponta CasADi, amplamente discutido em termos do estado-da-arte, c.f. \citep{andersson2019casadi,gros2020linear}. Nas Figuras \ref{estados} e \ref{saida1}, apresentamos, respectivamente, a evolução dos estados do sistema, os sinais de controle obtidos com os dois algoritmos e as saídas controladas ($P_{fp}$ e $w_{g,in}$). Discorremos sobre resultados obtidos a seguir:
\begin{itemize}
    \item Primeiramente, enfatizamos que método LPV é capaz de obter um desempenho de controle numericamente equivalente ao obtido com o NMPC. Este fato é de grande relevância, uma vez que o software CasADi é usado, tanto na literatura tal como na prática industrial, como referência para algoritmos NMPC embarcados;
    \item Ademais, reforçamos que o esforço numérico médio necessário para o cálculo do sinal de controle, com a metodologia LPV, é muito reduzido, uma vez que apenas um QP é resolvido por amostra. Na Tabela \ref{performanceresultstable} sintetizamos este resultado, apresentando o tempo médio de cômputo da lei de controle com cada abordagem de controle ($t_c$), o desvio padrão correspondente ($\sigma(t_c)$);
    \item Em suma, podemos concluir que, de fato, a abordagem LPV MPC permite um desempenho de controle equiparável, com um cômputo tolerável para aplicações em tempo-real, inferior ao período de amostragem. Ressaltamos, também, que o tempo médio necessário para o cômputo, é dez vezes menor do que quando implementa-se um NMPC com CasADi.
\end{itemize}

\begin{table}[ht]
    \centering
      \caption{Comparação da desempenho.}
    \label{performanceresultstable}
    \small
    \begin{tabular}{c c c}
        \hline
         Metódo  &  $t_c$ & $\sigma (t_c)$ \\\hline 
         NMPC (CasADi) & $288$ ms   &$49.4$ ms \\ 
         LPV MPC & $26$ ms  &$20$ ms \\ \hline 
    \end{tabular}
\end{table}

\section{Conclusões}
\label{secconc}

Neste artigo, apresentamos um tutorial detalhado sobre o projeto de controladores preditivos não lineares com base em modelos LPV. Abordamos todas as etapas necessárias para a síntese, desde modelagem até detalhes de implementação. Ilustramos, inclusive, aspectos teóricos referentes à estabilidade em malha-fechada e à factibilidade recursiva da otimização. Por fim, apresentamos um estudo de caso numérico, comparando as abordagem MPC LPV e não linear, considerando um sistema de extração de petróleo por elevação artificial com injeção de gás. Ressaltamos as principais vantagens: 
\begin{itemize}
    \item A representação LPV permite uma descrição exata de dinâmicas não lineares, sem requerer qualquer tipo de aproximação ou linearização local; 
    \item A predição de dinâmicas futuras, com base em modelo LPV, é uma operação linear a cada amostra. Desta forma, o programa de otimização resultante torna-se computacionalmente mais suave que no caso não linear;
    \item Diversos resultados, tal como o apresentado neste artigo, indicam a similaridade de desempenho de controle das técnicas MPC LPV com algoritmos MPC não lineares, além da competitividade numérica quando comparadas a ferramentas MPC não lineares rápidas, tais como ACADO e CasADi. No estudo de caso do \textit{gas-lift}, a abordagem LPV MPC permite um tempo de cômputo onze vezes menor.
\end{itemize}

\section*{Agradecimentos}

\noindent Agradecemos: CAPES, CNPq ($304032/2019-0$, $403949/2021-1$) e o o apoio financeiro do Programa de Recursos Humanos da Agência Nacional do Petróleo, Gás Natural e Biocombustíveis - PRH-ANP, suportado com recursos provenientes do investimento de empresas petrolíferas qualificadas na Cláusula de P, D\&I da Resolução ANP n$^o$ $50$/$2015$.
\bibliographystyle{model5-names}
\bibliography{refssbai23tutorailLPVMPC}

\end{document}